# Third order nonlinear transport properties in topological chiral antiferromagnetic semimetal CoNb$_3$S$_6$

Junjian Mi[1,2,#], Jialin Li[2,3,#], Miaocong Li[1], Sheng Xu[1], Shuang Yu[1], Zheng Li[1], Xinyi Fan[2], Huanfeng Zhu[2,4], Qian Tao[1], Linjun Li[2,*], Zhuan Xu[1,5,6,*]

1. Zhejiang Province Key Laboratory of Quantum Technology and Device, School of Physics, Zhejiang University, Hangzhou 310027, China
2. State Key Laboratory of Modern Optical Instrumentation College of Optical Science and Engineering Zhejiang University, Hangzhou 310027, China
3. Hangzhou Institute for Advanced Study, University of Chinese Academy of Sciences, Hangzhou 310024, China
4. Jiaxing Key Laboratory of Photonic Sensing & Intelligent Imaging Jiaxing Institute Zhejiang University, Jiaxing 314000, China
5. State Key Laboratory of Silicon and Advanced Semiconductor Materials, Zhejiang University, Hangzhou 310027, China
6. Hefei National Laboratory, Hefei 230088, China

[#]These authors contributed equally to this work

*corresponding author. E-mail: lilinjun@zju.edu.cn; zhuan@zju.edu.cn

## Abstract

The topology between Bloch states in reciprocal space has attracted tremendous attention in recent years. The quantum geometry of the band structure is composed of quantum metric as real part and berry curvature as imaginary part. While the Berry curvature, the Berry curvature dipole and Berry connection polarizability have been recently revealed by the first order anomalous hall, second order and third order nonlinear Hall effect respectively, the quantum metric induced second order nonlinear transverse and longitudinal response in topological antiferromagnetic material MnBi$_2$Te$_4$ was only very recently reported. Here we demonstrate the similar third order nonlinear transport properties in the topological antiferromagnetic CoNb$_3$S$_6$. We observed that the third order nonlinear longitudinal $V_{xx}^{3\omega}$ increase significantly at the antiferromagnetic transition temperature $T_N$

~ 29 K, which was probably induced by the quantum metric without time-reversal symmetry or inversion symmetry. Besides, temperature-dependent nonlinear behaviour was observed in the first order I-V curve below the Neel temperature $T_N$, which was not reported in $MnBi_2Te_4$ and FeSn. Such nonlinear I-V behaviour hints for the possible existence of Charge Density Wave (CDW) state, which has been discovered in its sister material $FeNb_3S_6$. Simultaneously, two plateaus in the third order nonlinear longitudinal $V_{xx}^{3\omega} \sim I^\omega$ curve are observed, which is also speculated to be related with the possible CDW state. However, the genuine mechanism for the first order nonlinear I-V and its relation with the third order nonlinear transport call for more experimental investigations and theoretical interpretation. Our work provides a way to explore third harmonic nonlinear transport and interaction with magnetic order and CDW.

## Introduction

The intrinsic nonlinear hall responses are connected to topological properties of material, which have attracted tremendous interests in condensed materials physics in recent years[1, 2, 3, 4, 5, 6, 7, 8, 9, 10, 11, 12, 13, 14, 15, 16, 17]. The Berry curvature and the quantum metric are the imaginal part and real part of the quantum geometric tensor, respectively. They and their differential derivatives have been unveiled to be responsible to the interesting nonlinear transport phenomenon. For instance, the Berry curvature induces the anomalous Hall effect (AHE)[18]. The Berry curvature dipole can lead to second order nonlinear Hall effect and the Berry connection polarizability can induce third order nonlinear Hall effect in non-magnetic topological materials[1, 2, 19, 20, 21]. Recently, such second order nonlinear transport induced by the quantum metric dipole has been theoretically predicted and experimentally verified in topological antiferromagnetic material MnBi$_2$Te$_4$[22, 23, 24, 25]. It is worth noting that, the Berry curvature can only contribute to the transverse conduction in nonlinear transport. However, the quantum metric can contribute to the both longitudinal and transverse nonlinear conductivity. While the intrinsic contribution to the nonlinear transport is determined by the quantum geometric property, the extrinsic mechanism like Skew scattering and Side jump could also lead to nonlinear transport[5, 10]. Therefore, it is interesting to explore nonlinear transport in topological material, especially to discover its intrinsic quantum geometric contribution and cautiously exclude the possible extrinsic effect.

Third order nonlinear hall effect induced by the Berry connection polarizability tensor have been observed in nonmagnetic materials T$_d$-MoTe$_2$ and TaIrTe$_4$ [19, 20]. Recent theoretical work predicted Berry connection polarizability tensor can survive in certain magnetic topological material without time-reversal symmetry[26, 27]. At the same time, we noticed that Ref[28, 29] reported the quantum metric and Berry curvature induced third-order nonlinear transport phenomenon in topological antiferromagnetic material MnBi$_2$Te$_4$ and FeSn. Therefore, it is interesting to experimentally investigate the third nonlinear transport and their relation to the magnetic order. Therefore, we explore the third-order nonlinear transport in a topological antiferromagnetic semimetal CoNb$_3$S$_6$, which has a previously theoretically predicted Berry connection polarizability tensor.

The large AHE and anomalous Nernst effect have been reported recently in CoNb$_3$S$_6$[30, 31, 32], which are probably induced by its intrinsic Berry curvature. The Weyl points near the Fermi level were revealed by the Angle-resolved photoemission spectroscopy (ARPES) results and first-principle calculations were proposed for the cause of the large AHE[33,34]. Besides, the all-in-all-out type non-coplanar antiferromagnetic (AFM) order have been proposed through polarized neutron scattering experiments[35]. These interesting phenomena indicate $CoNb_3S_6$ is an attracting platform to investigate the interaction between its nonlinear transport properties and magnetic order.

In this work, we measured the third order nonlinear longitudinal $V_{xx}^{3\omega}$ and transverse $V_{xy}^{3\omega}$ signals driven by the a. c. current with frequency $\omega = 13.33$ Hz in high quality exfoliated $CoNb_3S_6$ single crystal. With the decrease of temperature, the third order nonlinear longitudinal $V_{xx}^{3\omega}$ increases significantly near the antiferromagnetic transition temperature ($T_N \sim 29$ K), and the $V_{xx}^{3\omega}$ decreases with further decrease of temperature. The sign of transverse $V_{xy}^{3\omega}$ has a change with part of contribution from longitudinal $V_{xx}^{3\omega}$ due to the asymmetry of the electrode. Surprisingly, the first order I-V curve behaves nonlinear characteristic below the Neel temperature $T_N$. Such first order nonlinear I-V behaviour possibly corresponds to a Charge Density Wave (CDW) gap along with the antiferromagnetic order transition, which bears similarity to its sister material $FeNb_3S_6$[36]. Correspondingly, the third order nonlinear longitudinal $V_{xx}^{3\omega}$ -T curve has two plateaus with the amplitude of current increases which was not observed in $MnBi_2Te_4$ or other materials. The magnetic field-dependent third order nonlinear transverse $V_{xy}^{3\omega}$ undergoes a loop which is consistent with its first order anomalous hall effect. Such third order nonlinear anomalous hall effect and longitudinal behaviour below the Neel temperature $T_N \backsim$ 29 K are probably induced by the Berry curvature and Quantum metric respectively, as uncovered in systems without inversion symmetry and time-reversal symmetry, such as $MnBi_2Te_4$ very recently[29].

## Results

1. The significantly enhanced longitudinal third-harmonic nonlinear voltage $V_{xx}^{3\omega}$ at AFM transitional temperature:

The crystal structure of $CoNb_3S_6$ is presented in Fig. 1(a) with the space group of $P6_322$ where the Co atoms are inserted between the layers of $2H-NbS_2$. Fig. 1(b) shows the temperature dependence of magnetic susceptibility, measured under an applied magnetic field of 0.1T along c-direction, with field cooling (black line) and zero field cooling (read line). Antiferromagnetic transition $T_N \sim 29$ K is determined from the temperature dependence of magnetic susceptibility, which is consistent with the former experiment report[31] . The inset displays the optical image of single crystal $CoNb_3S_6$ which was synthesized by chemical vapor transport method as described in Methods. Fig. 1(c) and Fig. 1(d) show the temperature dependence of the longitudinal first harmonic $V_{xx}^{3\omega}$ voltage and third-harmonic voltage $V_{xx}^{3\omega}$ from 2 K to 300 K with applied a. c. current of 1.80 mA with the frequency f = 13.33 Hz. The temperature dependence of the longitudinal first-harmonic shows a metallic behaviour with a kink near 28 K ($T_N$), which is smaller than the AFM transition $T_N \sim 29$ K in magnetic susceptibility measurements. The decreasing of AFM transition temperature of $CoNb_3S_6$ in Fig.1c is probably restrained by the large current which was further studied with different amplitude of a. c. current as shown in Fig .4. The inset shows the atomic force microscopy image of the device. Fig. 1(d) shows the temperature dependence of the third-harmonic voltage $V_{xx}^{3\omega}$ from 2 K to 300 K. A peak third harmonic voltage $V_{xx}^{3\omega}$ of 0.92 mV was obtained at 26.6 K. Inset shows the detailed longitudinal third harmonic voltage $V_{xx}^{3\omega}$ from 2 K to 40 K.

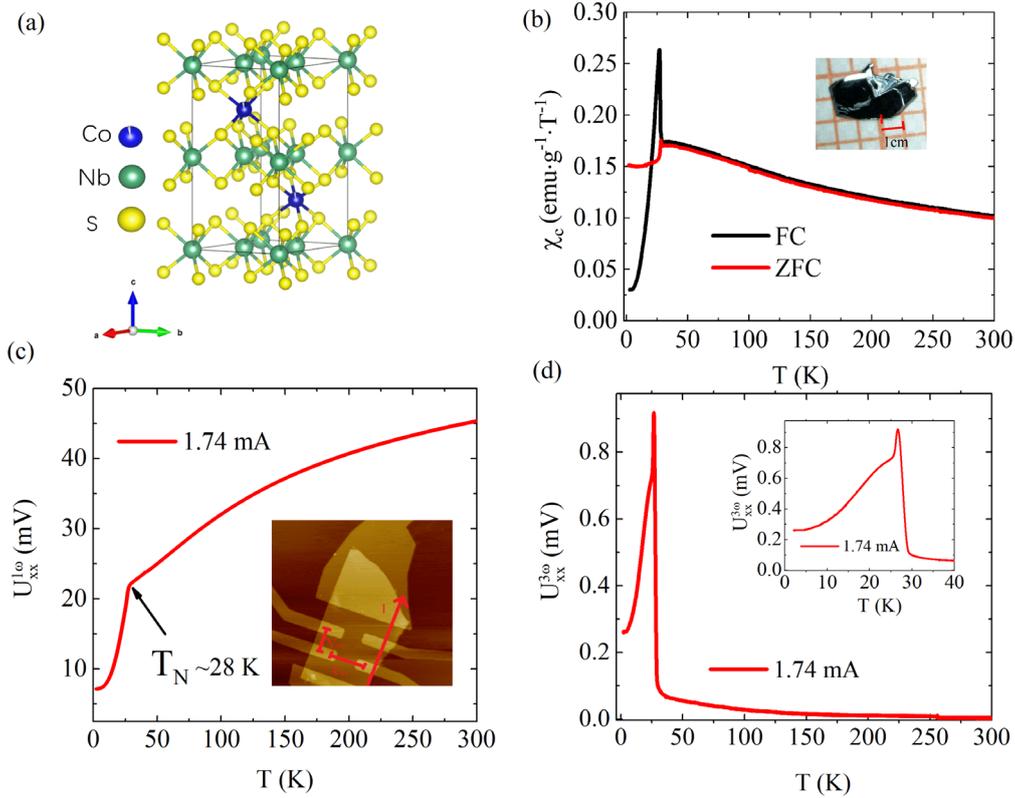

**Fig. 1. Crystal structure, magnetic and electric transport properties of CoNb$_3$S$_6$.** (a) The crystal structure of CoNb$_3$S$_6$. (b) Magnetic susceptibility measured with magnetic field along c-axis. Inset shows the optical image of as grown single crystal CoNb$_3$S$_6$. (c) Temperature dependence of the longitudinal first harmonic voltage $V_{xx}^{1\omega}$ from 2 K to 300 K with applied a. c. current of 1.74 mA. Inset shows the atomic force microscopy image of the device used for electrical transport measurements, the sample thickness is about 100 nm. (d) Temperature dependence of the longitudinal third harmonic voltage $V_{xx}^{3\omega}$ from 2 K to 300 K with applied a. c. current of 1.74 mA.

Inset shows the longitudinal third harmonic voltage $V_{xx}^{3\omega}$ from 2 K to 40 K.

## 2. The magnetic field-dependent nonlinear transport properties in CoNb$_3$S$_6$.

Fig. 2(a) and Fig. 2(b) show the magnetic field dependence of the longitudinal first harmonic voltage $V_{xx}^{3\omega}$ and transverse first harmonic voltage $V_{xy}^{3\omega}$ respectively at different temperatures below T$_N$ with applied a. c. current of 1.74 mA and frequency f = 13.33 Hz. The longitudinal $V_{xx}^{3\omega}$ exhibits even symmetry while the transverse $V_{xy}^{3\omega}$ exhibits odd symmetry under positive and negative magnetic fields, which is consistent with the former reports. The magnetic hysteresis phenomenon is originated from its ferromagnetic order under out-of-plane high magnetic field below the T$_N$. However, the amplitude of third harmonic voltage $V_{xx}^{3\omega}$ decreases as the magnetic field increases, as shown in Fig. 2(c), which is different from first harmonic voltage. Fig. 2(d) shows the odd symmetry properties of $V_{xx}^{3\omega}$ change with magnetic field, and the coercive field increases as the temperature decreases. The longitudinal harmonic voltage $V_{xx}^{n\omega}$ (B) was calculated by the formula $V_{xx}^{n\omega}$ (B) = ($V_{xx}^{n\omega}$ ' (B) + $V_{xx}^{n\omega}$ ' (-B)) / 2 and transverse harmonic voltage $V_{xy}^{n\omega}$ (B) = ($V_{xy}^{n\omega}$ '(B) - $V_{xy}^{n\omega}$ '(-B)) / 2 to exclude the influence of the electrode asymmetry. The initial data of harmonic voltage change with magnetic field is shown in SI.1.

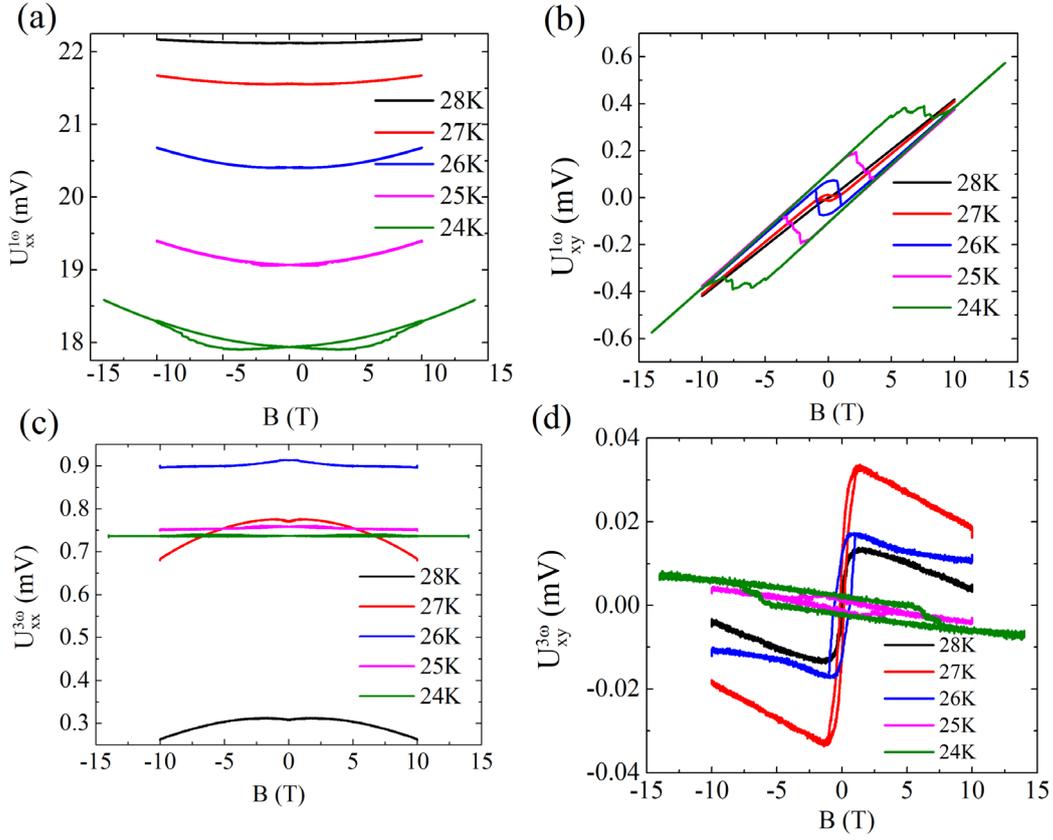

**Fig.2. The magnetic field-dependent nonlinear transport properties in CoNb$_3$S$_6$.** (a-d) The out-of-plane B-fields dependence of the longitudinal first-harmonic voltage $V_{xx}^{1\omega}$ (a), transverse first harmonic voltage $V_{xx}^{1\omega}$ (b), third-harmonic voltage $V_{xy}^{3\omega}$ (a), transverse third harmonic voltage $V_{xy}^{3\omega}$ (b), measured at different temperatures.

3. **The a.c. current amplitude-dependent longitudinal transport properties in CoNb$_3$S$_6$.**

Fig. 3(a) shows the a. c. current amplitude-dependent longitudinal first harmonic transport properties at different temperatures in CoNb$_3$S$_6$. The nonlinear first harmonic I-V behave was observed below the AFM transition T$_N$ ~ 29 K as shown in Fig. 3(a). There was no hysteresis for I-V curves sweeping forward and backward, thus can exclude the influence of thermal effect (Fig. S2). Interestingly, the longitudinal third harmonic voltage $V_{xx}^{3\omega}$ varies with the amplitude of the a. c. current, accompany by a strange transition, as shown in Figure 3(b).

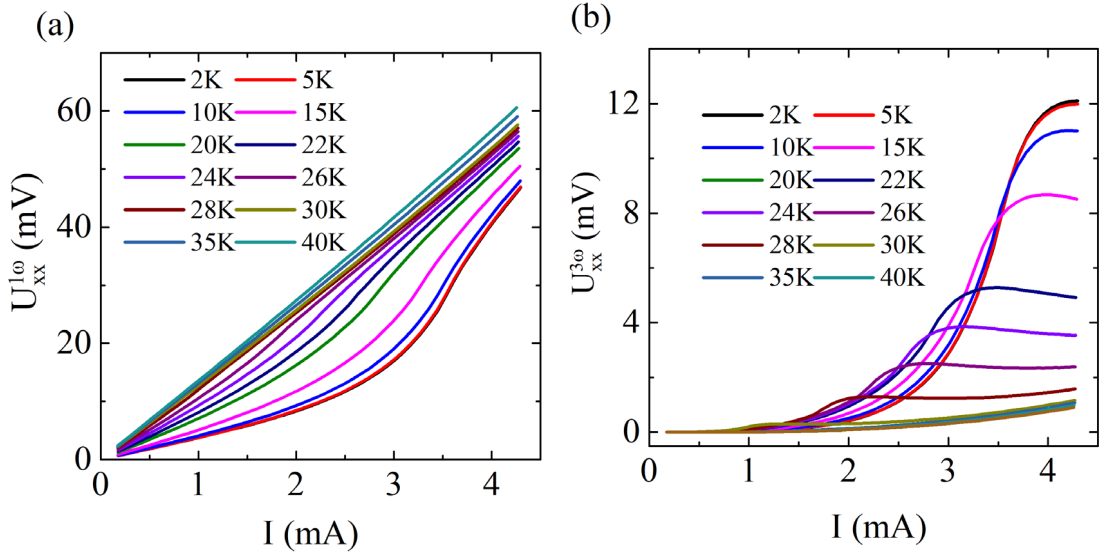

**Fig. 3. The a. c. current amplitude-dependent longitudinal transport properties in CoNb₃S₆.** (a-b) The amplitude of a. c. current dependence of the longitudinal first harmonic voltage $V_{xy}^{1\omega}$ (a) and third-harmonic voltage $V_{xy}^{3\omega}$ (b) at different temperature with zero magnetic field.

4. The temperature-dependent longitudinal nonlinear transport properties in CoNb₃S₆.

Fig. 4(a) shows the temperature-dependent longitudinal first-harmonic transport properties with different amplitude of a. c. current in CoNb₃S₆. As we can see, the transition temperature ($T_N$) decreases and the curves of $V_{xx}^{1\omega}$ - T become smoothly as the amplitude of a. c. current increases. The amplitude of longitudinal first harmonic voltage $V_{xx}^{1\omega}$ increases as the amplitude of a. c. current increases, which is consistent with the results of third-harmonic voltage $V_{xx}^{3\omega}$ shown in Fig. 3(b). Interestingly, the third-harmonic voltage $V_{xx}^{3\omega}$ doesn't show an obvious deceasing behaviour during the cooling process under high a. c. current amplitude (i.e. 4.35 mA). And the $V_{xx}^{3\omega}$ can reach 12 mV as the temperature decreases to 2 K with a. c. current of 4.35 mA.

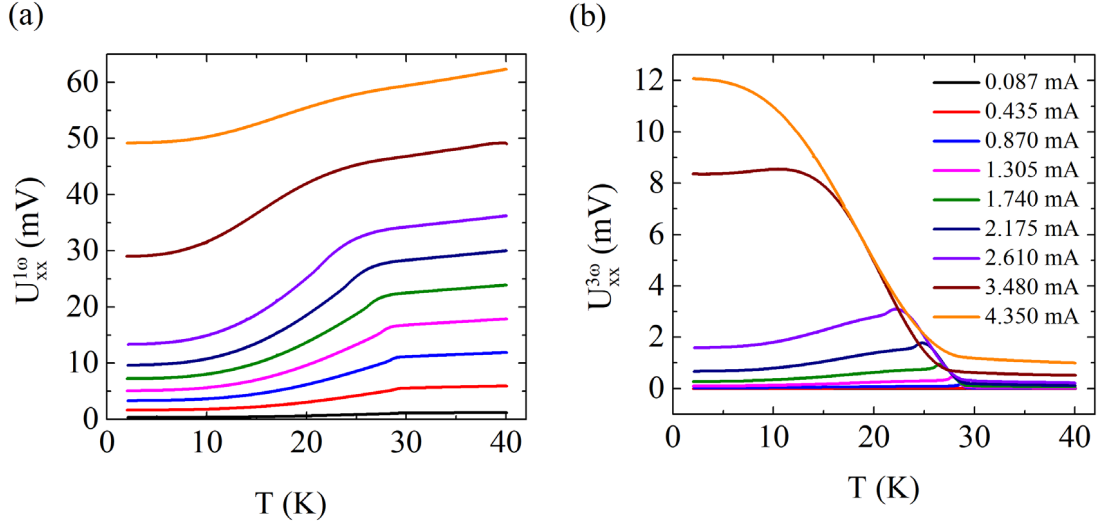

**Fig.4. The temperature-dependent longitudinal nonlinear transport properties in CoNb$_3$S$_6$.** (a-b) Temperature dependence of the $V_{xy}^{1\omega}$ (a) and $V_{xy}^{3\omega}$ (b) at different amplitude of a. c. current with zero magnetic field.

## Discussion

The third harmonic nonlinear transport was largely enhanced at AFM transition temperature which is probably connected to its magnetic quadrupoles or the quantum geometrical quadrupoles. The magnetic field dependence of even symmetry $V_{xx}^{3\omega}$ and odd symmetry $V_{xy}^{3\omega}$ at different temperature probably induced by the quantum geometry quadrupoles effect, skew scatting or side jump effect. The origin of the third harmonic nonlinear transport needs further calculations. Besides, the nonlinear first harmonic I-V behaviour have been observed below the AFM transition (T$_N$ ~ 29 K) which was probably induced by CDW transition, which is similar as its sister material FeNb$_3$S$_6$. The AFM transition temperature (T$_N$) decreases as the amplitude of a. c. current increases and the curves of $V_{xx}^{1\omega}$ - T become smoothly at AFM transition, which the large current suppress the AFM order probably. Also, the third harmonic transport is influenced by the similar effect.

## Conclusion

In summary, we measured the third order nonlinear longitudinal $V_{xx}^{3\omega}$ and transverse $V_{xy}^{3\omega}$ in topological chiral antiferromagnetic semimetal $CoNb_3S_6$. We observed the third order nonlinear longitudinal $V_{xx}^{3\omega}$ increases significantly as the temperature decreases near the AFM transition temperature $T_N$. Besides, the third harmonic nonlinear transverse $V_{xy}^{3\omega}$ varied with magnetic field with a loop which is consistent with its first harmonic anomalous hall effect. The magnetic field dependence of even symmetry $V_{xx}^{3\omega}$ and odd symmetry $V_{xy}^{3\omega}$ under different temperature probably induced by the quantum geometry quadrupoles effect. The first order I-V curve behave nonlinear characteristic was observed below the Neel temperature $T_N$, which is possibly corresponded to a gap which induced by the possible CDW order with an antiferromagnetic order transition. At the same time, the third harmonic nonlinear longitudinal $V_{xx}^{3\omega}$ -T curve has two plateau as the amplitude of current increases, which is connected to the CDW state. Our study opens the door to explore the origin of third harmonic nonlinear and interaction with AFM and CDW orders.

## Methods

**Materials Synthesis**

CoNb$_3$S$_6$ single crystals of were synthesized by chemical vapor transport (CVT) method. Firstly, polycrystalline CoNb$_3$S$_6$ was prepared by heating the mixture of Co, Nb and S powder with a stoichiometric of 1:3:6 at 900 °C for 5 days. Then, 2 g of the powder and 0.75 g/mL iodine were sealed in a quartz tube to grow single crystals with a temperature gradient 950 °C to 850 °C for 10 days. The single-crystal x-ray diffraction (XRD) performed on the PANalytical x-ray diffractometer with a Cu K$_{\alpha1}$. The atomic composition of single crystal CoNb$_3$S$_6$ was checked by the dispersive x-ray spectroscopy (EDS).

**Device fabrication**

The thin flakes were obtained by the traditional Scotch tape exfoliation method. The hall bar-shaped electrodes were fabricated by standard electron beam lithography and metal deposition (5nm Cr/ 100 nm Au). The thickness of the thin samples was determined by the atomic force microscopy.

**Magnetic and transport property measurements**

The magnetic susceptibility is measured on the Quantum Design Magnetic Properties Measurement System (MPMS - 5 T). The transport measurements are performed on a Quantum Design physical property measurement system (QD PPMS-14 T). The a. c. voltage measurements were performed with SR830 using the standard four-point method.

# Data Availability

For The data that support the findings of this study are available from the corresponding author upon reasonable request.

## Acknowledgements

Z.A Xu acknowledges the funding from National Key R&D Program of China (2019YFA0308602) and National Science Foundation of China (11774305), the Innovation program for Quantum Science and Technology (Grant No. 2021ZD0302500), L.J Li acknowledges the funding from National Key R&D Program of China (2019YFA0308602), National Science Foundation of China (11774308) and the Zhejiang Provincial Natural Science Foundation of China (LR20A040002).

## Contributions

J.J Mi and J.L Li contributed equally to the work. J.J Mi and S Xu performed the crystal growth,